\documentclass[amsmath,prl,aps,twocolumn]{revtex4}
\usepackage{amsthm,amsfonts,graphicx,verbatim}

\newcommand{\Tr}{{\rm Tr}}

\newcommand{\D}{{\rm d}}

\newcommand{\be}{\begin{equation}}
\newcommand{\ee}{\end{equation}}
\newcommand{\bea}{\begin{eqnarray}}
\newcommand{\eea}{\end{eqnarray}}

\newcommand{\la}{\langle}
\newcommand{\ra}{\rangle}
\newcommand{\lb}{\left[}
\newcommand{\rb}{\right]}
\newcommand{\lp}{\left(}
\newcommand{\rp}{\right)}
\renewcommand{\vec}[1]{{\bf #1}}
\renewcommand{\epsilon}{\varepsilon}

\begin{document}

\title{Scaling of entanglement entropy and superselection rules}
\author{Israel Klich$^{1,2}$ and Leonid Levitov$^{1,3}$ }
\affiliation{${}^1$ Kavli Institute for Theoretical Physics,
University of California Santa Barbara, CA
93106 \\
 ${}^2$
Department of Physics, University of Virginia, Charlottesville, VA 22904
\\
${}^3$ Department of Physics, Massachusetts Institute of Technology, Cambridge MA 02139}


\begin{abstract}
Particle number conservation in fermionic systems restricts the
allowed local operations on bi-partite systems. We show how this
restriction is related to measurement entropy of particle
fluctuations and compute it for
several regimes of practical relevance.
The accessible entanglement entropy restricted by
particle number conservation is equal, to leading order, to the full entanglement entropy.
The correction is bounded by the
log of the variance of particle number fluctuations.
The results are applied to generic Fermi systems in dimension $d>1$ as well as to several
critical systems in $d=1$.
\end{abstract}

 \maketitle

Entanglement entropy is a quantity which is often used in quantum
information theory to characterize the degree of entanglement
between different parts  of a quantum system that share the same
microscopic variables \cite{Bennet entropy}. Prior to its adoption
in the field of quantum information, entanglement entropy has been
introduced and studied in field theory as a possible contribution
to the entropy of black holes
\cite{Bombelli86,Srednicki93,Holzhey94}, where it was called
``geometric entropy.'' Recently it emerged as a quantity of
interest in many-body theory, where it was employed as a measure
of large-scale, nonlocal correlations signaling criticality
\cite{VidalLatorreetal} (for a review of these developments see
Ref.\cite{Amico} and references therein).


The non-local correlations between entangled quantum variables are
revealed by measurements made locally on two (or more) subsystems
of a quantum system. The locality of observables used to detect
entanglement is of particular importance in applications such as
teleportation \cite{Bennet teleportation} and quantum cryptography
\cite{Ekert}, which require that the two subsystems do not
interact after the initial state is prepared. However, our ability
to probe entanglement by a combination of local measurements is
often constrained by conservation laws (e.g. of particle number,
charge or spin), rendering some local observables physically
inaccessible.


Constraints on the local operations (known as super-selection
rules) can limit experimentally accessible entanglement. The
interplay of super-selection and entanglement was first analyzed
in Ref.\cite{Wiseman Vaccaro}, where the
accessible entanglement was quantified by averaging of
entanglement entropy over super-selection sectors. The quantity
defined in Ref.\cite{Wiseman Vaccaro}, which we refer to as
\emph{accessible entanglement entropy} was further discussed in
Refs.\cite{Banuls,Dowling, Beenakker}. A different point of view
on accessible entanglement was put forward in Ref.\cite{Verstraete
Cirac}, where super-selection constraints on local operations are
treated as a resource that can be employed for hiding information
in correlations which are blocked from local probing.


To understand the general relation between accessible entanglement
entropy and the full entanglement entropy, in this Letter we
present general results for fermion systems, interacting or
noninteracting. Fermion systems play a special role in the theory
of entanglement entropy because, on one hand, they are many-body
systems exhibiting rich and interesting behavior and, on the other
hand, they provide a good model of experimentally relevant
settings. In particular, the scaling of the full entanglement
entropy in Fermi systems depends on the nature of the state. If
the system is gapped, the entropy scales with the area of the
boundary of the region, $S\propto L^{d-1}$, where $L$ is the
linear size of the region ({\it c.f.} \cite{Plenio Eisert,Wolf
Verstraete}). However if the system is in a gapless (metallic)
state, the entropy scales as $S\propto L^{d-1}\log L$
\cite{Wolf,Gioev Klich}.
Furthermore, dynamics of fermionic systems can be used to generate complex entangled states. In particular, generation of entanglement in mesoscopic
conductors occurs naturally as a result of elastic scattering of
electrons on barriers and disorder potential \cite{Beenakker03,
Samuelsson}. A scheme for detecting electronic entanglement which
can discriminate between occupation number entanglement and mode
entanglement has been proposed in \cite{Giovannetti}.


One particularly attractive aspect of Fermi systems is that the
full many body entanglement entropy of free fermions can be linked
to experimentally accessible quantities, such as particle number
fluctuations \cite{KlichRefaelSilva} or current fluctuations
\cite{Klich Levitov}. The latter quantity was discussed in detail
in a recent proposal \cite{Klich Levitov} of an experiment to
measure the entanglement entropy by detecting electric noise
generated in a process of connecting two Fermi
seas through a quantum point contact (QPC).


While proposals such as \cite{KlichRefaelSilva,Klich Levitov} as
well as studies of scaling of entropy in fermion systems
\cite{Wolf,Gioev Klich} are of interest from the quantum
information perspective, they do not make a distinction between
entanglement entropy and the accessible entanglement entropy.
Given that only the accessible entropy can be used as a source of
entanglement for quantum information applications, here we set out
to investigate the effect of super-selection rules on the scaling
of the accessible entanglement in many-body systems of fermions.
We find that, under very general assumptions,
super-selection rules result in a difference between the
accessible and full entanglement entropies which is small in a
relative sense, provided that the number of Fermi particles
contributing to the entropy is large, $N\gg1$. It means that, in
essence, the schemes proposed for measuring the full entanglement
entropy can also yield a good estimate of the accessible
entanglement entropy, and vice versa.


To state our results in a quantitative form,  let us recall that
entanglement entropy is defined for a system partitioned into two
parts $A$ and $B$. The quantum state of the system is projected on
$A$, giving the reduced density matrix $\rho_A=\Tr_{B}\rho$, where
all degrees of freedom outside $A$ have been integrated out.
Entanglement entropy is then given by the von Neumann formula:
$S_A=-\Tr\rho_A\log\rho_A$. When super-selection rules are
present, a natural quantity to consider is $S_{A}^{\rm res}$
\cite{Wiseman Vaccaro}, which is obtained by averaging
entanglement entropy over subspaces with fixed conserved quantity
(in our case, particle number) as will be described below. We show
that if the state $\rho$ has a fixed number of particles then
\begin{eqnarray}\label{main inequality}
S_A-\Delta S
\leq S_{A}^{\rm res}\leq S_A
,\quad \Delta S = {1\over 2}\log\lb 2\pi e (C_2+{\textstyle{1\over 12}})\rb
\end{eqnarray}
where $C_2=\la (N_A-\la N_A\ra)^2\ra$ is the variance in the number of particles in
subsystem $A$. An immediate consequence is that since typically $C_2$ can not
 grow faster than
polynomially with the size of the subregion, the correction
$\Delta S$ to the entropy due to particle conservation is at most
logarithmic in the volume.

In many-body systems, where entropy scaling is of interest,
entanglement entropy $S_A$ of translationally invariant systems
(gapless as well as gapped) in space dimension $d>1$ typically
scales at least as the boundary area $L^{d-1}$, where $L$ is the
size of the region $A$. Given that $\Delta S$ in Eq.\eqref{main
inequality} grows at most as $\log L$, we see that the correction
to the entropy from the terms violating super-selection is
sub-leading to the entropy $S_A$. The situation may be more
complicated in dimension $d=1$, where for {\it critical systems},
described by a conformal field theory, entropy of a region of
length $L$ scales as $\log L$ \cite{Cardy}. Still, even in this
case, if the fermion density-density correlations decay as $1/r^2$ or faster,
the quantity $\log C_2$ grows at most as $\log \log
L$, and thus $\Delta S$ is again sub-leading to $S_A$. Several
systems of interest exhibiting this behavior are analyzed below.

Perhaps surprisingly, we found that $S_{A}^{\rm res}$ may be
computed explicitly in many experimentally relevant settings. This
is largely due to the fact that the difference $S_A-S_{A}^{\rm
res}$ is nothing but the measurement entropy $S_{\rm m}$ of particle number in
subregion $A$. We analyze the accessible entanglement entropy for
several cases: Entanglement generated  when two Fermi seas are
connected via a QPC for a time $\Delta t$ and then disconnected,
entanglement generated by a dc current in a QPC biased by voltage
$V$ during time $\Delta t$, entanglement in a Luttinger liquid and
in a d-dimensional free fermion system. The results for
$S_{\rm m}=S_A-S_{A}^{\rm res}$  are summarized in Table I.


\begin{table}[h]\label{table1}
\begin{tabular}{|c|c|}
  \hline
  {\rm System} & $S_{\rm m}=S_A-S_{A}^{\rm res}$ \\
  \hline
 {\rm  Fermi seas connected via a QPC (Fig.1a)}  &  $\log\log(\Delta t/\tau)$ \\
{\rm  DC voltage-biased QPC (Fig.1b)} & $\log(\Delta t eV /h)$ \\
{\rm Chiral Luttinger liquid}  & $\log\log(\tilde{L})$ \\
{\rm Fermions in dimension $d$}  &  $ \log(\tilde L^{d-1}\log \tilde L)$\\
  \hline
\end{tabular}
\caption{The difference between full and accessible entropy for
the systems described in the text ($\tau$ is a short time cutoff,
$\tilde L=k_FL$, where $k_F$ is Fermi momentum
and $L$ is system size). }
\end{table}

\begin{figure}
\includegraphics*[scale=0.4]{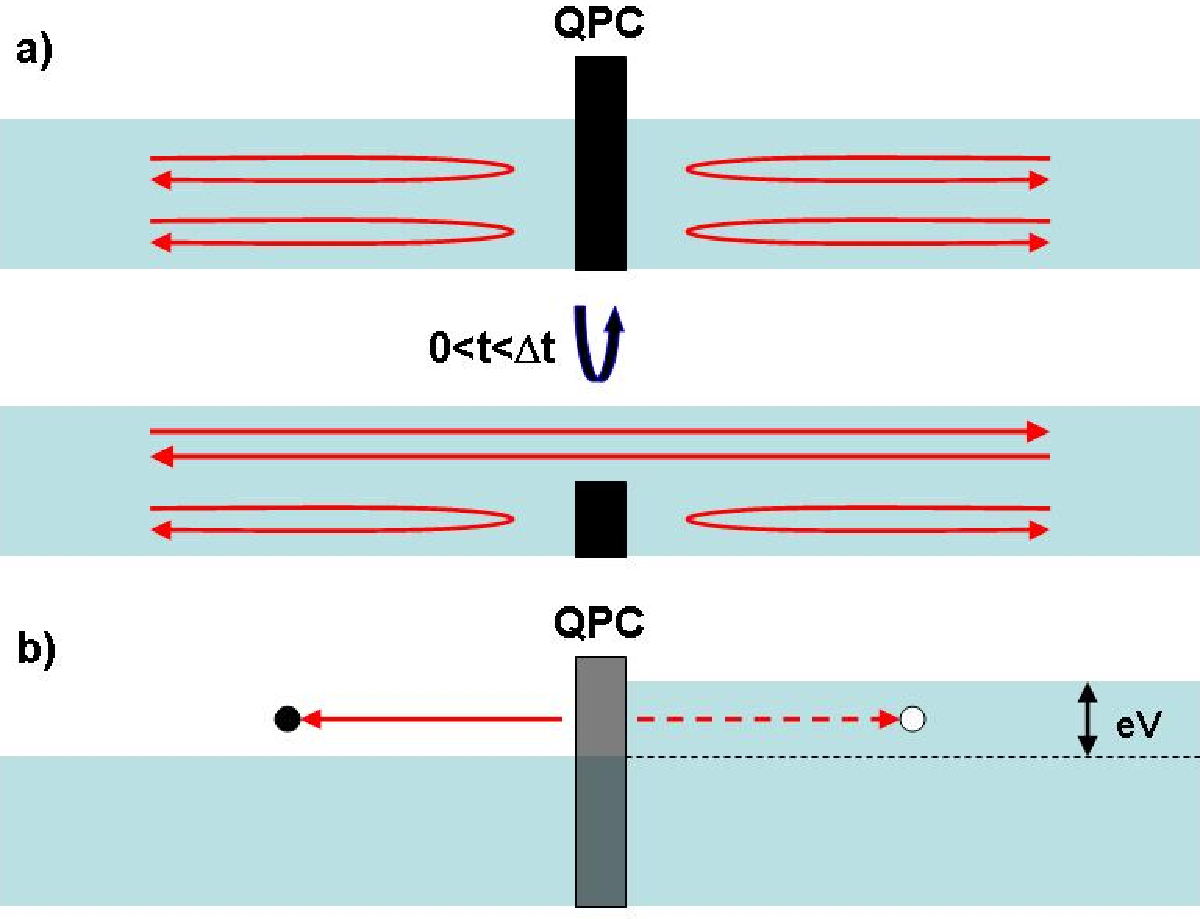}
\caption{ Schematic many-body evolution in a quantum point contact
(QPC), generating entanglement and current fluctuations. Two
fermionic reservoirs are coupled through a tunable tunnel barrier,
switching between fully closed and fully open states (a), or being
time-independent (b). In the case (a), the left and right
reservoirs, which are connected during $0<t<\Delta t$, and
disconnected at earlier and later times, are maintained at equal
chemical potentials. In the case (b) the QPC is biased by a DC
voltage $V$, with transmission fixed at a constant value $0<D<1$.
}
\label{fig1}
\end{figure}


\section{Superselection and Entanglement for Two Fermions}

We begin with some
general comments on the nature of entanglement of
fermions and the effect of particle number conservation.
To
motivate the definition we use for the accessible entanglement
entropy, we briefly
 discuss
what entanglement of fermions
means, and how
conservation
  laws
enter the picture.
In that, for reader's sake, we restate some of the observations made in Ref.\cite{Wiseman Vaccaro}.

Fermionic wavefunctions are always antisymmetrized, but this
property by itself does not imply entanglement. To illustrate the
role of antisymmetrization and of particle conservation we focus
on the simplest case of just two fermions. Consider a one
dimensional interval $0<x<L$, partitioned into   equal halves $A$
and $B$. We define $A(x)=1$ if $0<x<L/2$ and $A=0$ otherwise and
similarly $B(x)=1$ if $L/2<x<L$ an $B=0$ otherwise.


Let us compare the properties of two antisymmetric wavefunctions,
\begin{eqnarray}
\psi_1(x,y)={\sqrt{2}\over L}(A(x)B(y)-A(y)B(x))
\end{eqnarray}
and
\begin{eqnarray}
\psi_2(x,y)={1\over \sqrt{2}L}(e^{4\pi i x\over L}-e^{4\pi i
y\over L}).
\end{eqnarray}
%
The state $\psi_1$ is obtained by putting one fermion in box $A$ and another in box $B$.
While this state may appear ``entangled" due to the
antisymmetrization, it is of course not entangled. The reduced
density matrix in $B$ is just a projection on a single particle
with wavefunction $B(x)$, and thus it is a pure state. In other
words, lack of information about the state in $A$, does not
increase the entropy of the state in $B$. Thus, {\it
anti-symmetrization of the wavefunction by itself does not produce
entanglement}.


In contrast, the state $\psi_2$, obtained by putting one particle
in the state ${1\over \sqrt{L}}$ and another particle in the
state ${1\over \sqrt{L}}e^{4\pi i x\over L}$, contains
entanglement. To see this we decompose $\psi_2$ as follows:
\begin{eqnarray}
\psi_2(x,y)=F_{AA}+F_{BB}+F_{AB}+F_{BA}
\end{eqnarray}
where
\begin{eqnarray} &
F_{AA}={1\over{\sqrt{2}L}}(e^{4\pi i x\over L}-e^{4\pi i y\over
L})A(x)A(y)\\ & \nonumber F_{BB}={1\over{\sqrt{2}L}}(e^{4\pi i
x\over L}-e^{4\pi i y\over L})B(x)B(y)\\ & \nonumber
F_{AB}={1\over {\sqrt{2}L}}(e^{4\pi i x\over L}A(x)B(y)-e^{4\pi i
y\over L}A(y)B(x))\\ & \nonumber
F_{BA}={1\over{\sqrt{2}L}}(e^{4\pi i x\over L}B(x)A(y)-e^{4\pi i
y\over L}B(y)A(x))
\end{eqnarray}

The $F_{AA}$ and $F_{BB}$  wavefunctions represent the
possibility that both fermions are simultaneously in $A$ or in
$B$. Let us now measure the number of particles in $A$. If the
result of this measurement is $0$ or $2$, the wavefunction is
collapsed into $F_{AA}$ or $F_{BB}$. However, if we find just one
particle in $A$, we have collapsed the wavefunction into the
state $F_{AB}+F_{BA}$.

The wavefunction $F_{AB}+F_{BA}$ is maximally entangled, in
complete analogy with a singlet spin state of
Einstein-Podolsky-Rosen form. Indeed, we can denote
$|\uparrow\ra_A$ and $|\downarrow\ra_A$ as having a particle in
$A$ in the modes ${\sqrt{2 \over  L}}e^{4\pi i x\over L}A(x)$ and
${\sqrt{2\over L}}A(x)$, and similarly for
$|\uparrow\ra_B $ and $|\downarrow\ra_B$ (the form
$e^{4\pi ix\over L}$ was chosen so that
${}_A\la\uparrow|\downarrow\ra_A=0$).
In this notation we have
\begin{eqnarray}
F_{AB}+F_{BA} = {1\over
\sqrt{2}}(|\uparrow\ra_A|\downarrow\ra_B+|\downarrow\ra_A|\uparrow\ra_B)
\end{eqnarray}
Locally measuring whether the ``up" or ``down" mode is occupied, and
locally moving a particle between these states (for example by
coupling to a potential) are valid physical operations
corresponding in spin notation to $\sigma_z$ and $\sigma_x$.

This entanglement can be contrasted with the properties of the
two-particle state $F_{AA}+F_{BB}$. While the state
$F_{AA}+F_{BB}$ is formally entangled,
this entanglement can not be revealed by Bell measurements. The
states analogous to $\downarrow$ and $\uparrow$ here will be
``having no particles in A" and ``having two particles in A", which
could be also associated with spin variables $\sigma_z=\pm1$ and
directly measured.
However, because of particle number conservation, we are blocked
from rotating between these ``up" and ``down" states, and so
$\sigma_x$ is inaccessible. This blocks us from a necessary
ingredient to perform Bell measurement, rendering this
entanglement ``unuseful". We have thus seen that in this example,
with probability $1/2$, we can perform measurements that violate
Bell inequalities on $\psi_2$.

Consider a large number $n$ of systems with the wavefunction
$\psi_2$. Let $p$ denote the probability of measuring one charge
in $A$ in state $\psi_2$. If we now measure the number of
particles in $A$, we will average $pn$ cases of (maximally
entangled pairs) $F_{AB}+F_{BA}$.  This motivates defining the
entanglement of the system by averaging over the probabilities to
fall in different super-selection sectors.

The discussion above, was concerned with the case of copies of a
simple system. In contrast, when considering partitioning of many
body systems, we are interested in {\it scaling} properties. Here,
the entanglement content of a single system grows in a non-trivial
way as the system size is increased. Thus, we are considering the
entropy scaling in many-body systems, when super-selection rules
are applied.


\section{Derivation of the main result}

We now proceed to introduce the super-selection accessible
entanglement entropy. We assume that the local operations we may
perform in $A$ and $B$ respect conservation of quantities $\hat
N_A$ and $\hat N_B$. The super-selection sectors are conveniently
represented as the invariant subspaces of a an operator
$\hat{N}=\hat{N}_A\otimes I+I\otimes \hat{N}_B$ (where $I$ is the
identity operator). Thus $\hat{N}$ represents the globally
conserved quantity (i.e. total number of particles) and
$\hat{N}_A$ ($\hat{N}_B$) are the locally conserved quantities,
which can be thought of as ``number of particles in $A$ ($B$)". The
available entanglement entropy is obtained by averaging over the
entropy obtained by first restricting the density matrix to a
sector with fixed numbers of particles $N_A$, $N_B$ \cite{Wiseman
Vaccaro}. More concretely, this is done using the density matrices
%
\begin{eqnarray}\label{rho_n,m}
\rho_{n,m}={1\over p_{n,m}}\Pi^A_{n}\otimes \Pi^B_{m}\rho \Pi^A_{n}\otimes \Pi^B_{m}
,
\\
\label{p_n,m}
p_{n,m}=\Tr(\Pi^A_{n}\otimes \Pi^B_{m}\rho \Pi^A_{n}\otimes \Pi^B_{m} )
,
\end{eqnarray}
where $\Pi^A_{n}$ are projectors onto sectors with fixed particle
number $n$ in $A$ (i.e. all states $\psi$ in $A$, such that
$\hat{N}_A\psi=n\psi$) and similarly $\Pi^B_{m}$ projects on
sectors with $N_{B}=m$ in $B$.
%
%

Following Ref.\cite{Wiseman Vaccaro}, we define the accessible
entropy as
\begin{eqnarray}\label{AccessEntropy}
S_{A}^{\rm res}=-\sum p_{n,m} \Tr (\rho_{n,m})_A\log (\rho_{n,m})_A
\end{eqnarray}
where $(\rho_{n,m})_A=\Tr_B \rho_{n,m}$.  In terms of the
probabilities \eqref{p_n,m} we can write the Shannon entropy
\be\label{S_m_general} S_{\rm m}=-\sum_{n,m} p_{n,m}\log p_{n,m} ,
\ee which gives the {\it measurement} entropy of $\hat{N}_A$ and
$\hat{N}_B$\cite{Balian}.

We now turn to the proof of our main result \eqref{main
inequality}. We assume that we know that the state of the system
$|\psi\ra$ is characterized by a fixed total number of particles,
${\hat{N}}|\psi\ra=N|\psi\ra$. The Schmidt decomposition of the
state $|\psi\ra$ may be written as $|\psi\ra=\sum_n
C_n^{\alpha}|n,\alpha\ra_A\otimes|N-n,\alpha\ra_B$, where $\alpha$
enumerates different states with the same quantum number, which equals $n$ in
$A$, and $N-n$ in $B$. Tracing over $B$ we see that only the
reduced density matrices $(\rho_{n,m})_A$ with $m=N-n$ are
nonzero, giving
\begin{eqnarray}
S_{A}^{\rm res}=-\sum_{n} p_{n,N-n} \Tr (\rho_{n,N-n})_A\log (\rho_{n,N-n})_A
.
\end{eqnarray}
%
Using the constraint $n+m=N$, we can also simplify the expression
(\ref{rho_n,m}), writing it as
\begin{eqnarray}
 (\rho_{n,N-n})_A={1\over p_{n,N-n}}  \Pi^A_{n}\rho_A \Pi^A_{n}
.
\end{eqnarray}
Combining this formula with the representation $\rho_A=\sum_n
p_{n,N-n}(\rho_{n,N-n})_A$, and noting that
$(\rho_{n,N-n})_A(\rho_{n',N-n'})_A=0$ if $n\neq n'$, we can write
the accessible entropy \eqref{AccessEntropy} as
\begin{eqnarray}\label{Sr=S-Sm}
S_{A}^{\rm res} &=& 
-\sum_n
\Tr  {(\Pi_n \rho_A\Pi_n)}\log {\Pi_n \rho_A\Pi_n\over p_{n,N-n}}\\ \nonumber
&=& -\Tr \rho_A\log\rho_A+\sum_n p_{n,N-n}\log p_{n,N-n}\\ \nonumber
&=& S_A-S_{\rm m}
.
\end{eqnarray}
Here $S_{\rm m}=-\sum_n p_{n,N-n}\log p_{n,N-n}$ is the
measurement entropy \eqref{S_m_general} for fixed total particle
number $N=n+m$. The latter constraint makes particle number
measurements in $A$ and $B$ perfectly correlated, and so particle
fluctuations in $A$ alone are sufficient to determine $S_m$.

The relation
\eqref{Sr=S-Sm} may be interpreted as an extension of the well
known relation from information theory to the quantum case: If
$S(X,Y)$ is the Shannon entropy associated with the joint
probability distribution of the random variables $X,Y$, then
$S(X,Y)=S(X|Y)-S(Y)$ where $S(X|Y)$ is the entropy of $X$
conditioned on knowing $Y$, and $S(Y)$ is the entropy of $Y$.

Next, we estimate by how much $S_{A}^{\rm res}$ can depart from
$S_A$. This can be achieved using the relation $S_{A}^{\rm res} =
S_A-S_{\rm m}$ derived above, and maximizing $S_{\rm m}$. It is a
basic result of information theory that given a variance $C_2$ and
mean $C_1$, a \emph{continuous} distribution maximizing the
entropy is a Gaussian distribution (this can be easily proved
using Lagrange multipliers). The entropy of a Gaussian
distribution, $S_{0}={1\over 2}\log 2\pi e C_2$, where $C_2=\la\la
N_A^2 \ra\ra$ thus supplies an upper bound on measurement entropy
of a continuous observable. For a \emph{discrete} variable, such
as particle number, charge, or spin, this upper bound on entropy
has to be slightly modified \cite{Cover Thomas} to be
\be\label{S_m<Delta S}
S_{\rm m}\leq \Delta S \equiv {1\over 2}\log\lb 2\pi e (C_2+\textstyle{{1\over 12}})\rb
\ee
establishing the inequalities \eqref{main inequality}.


 It is interesting to  note that in Ref.\,\cite{klichJPA} the variance of particle
number of noninteracting fermions was shown to provide a lower
bound on the entanglement entropy, $S_A\geq(4\log2)C_2$. This
inequality can be used to write
\begin{eqnarray}\label{S difference for noninteracting}
S_A-S_{A}^{\rm res}\leq{1\over 2}\log\lb 2\pi e \lp{S_A\over 4\log 2}+{1\over 12}\rp\rb
\end{eqnarray}
which indicates that the difference $S_A-S_{A}^{\rm res}$ becomes
small in a relative sense as $S_A$ increases.

It is straightforward to generalize the result \eqref{main
inequality} to the case of several conserved quantities, which we
refer to as ``charges" $a_1$ ... $a_k$. For the measurement entropy
$S_{\rm m}(a_1...a_k)$ of
  these quantities,
using the subadditivity property of entropy, we have
\[
S_{m}(a_1...a_k)\leq
S_{m}(a_1)+...+S_{\rm m}(a_k)
.
\]
Using for each of the terms the inequality \eqref{S_m<Delta S}, we arrive at
\[
S_A-S_{A}^{\rm res} = S_{m}(a_1...a_k) \leq \sum_{i=1...k}{1\over
2}\log\lb 2\pi e \lp C_2(a_i)+\textstyle{1\over 12}\rp\rb ,
\]
where $C_2(a_i)$ is the variance in measurement of the charge $a_i$ ($i=1...k$).




\section{Measurement entropy for static and dynamical entanglement generation}

Now, having established the connection between accessible
entanglement entropy and measurement entropy, we turn to discuss
properties of the latter. In addition to being useful as a tool in
providing the above bounds on $S_A^{\rm res}$, the quantity
$S_{\rm m}$ is directly measurable, and thus is of interest in
itself.

  In general, of course,
the conserved charge does not have to be particle
number. For example, the conserved quantity may be chosen to be
energy. In this case, considered in Ref. \cite{Barankov}, it was argued that for
some situations of interest the measurement entropy of energy $S_{\rm m}$
can mimic the behavior of thermodynamic entropy
such as the second law  of thermodynamics,
and invariance under adiabatic evolution of the system.

A useful tool for computing $S_{\rm m}$ in many-body systems is
provided by the generating function, defined as a Fourier
transform of the probability distribution:
\begin{eqnarray}\label{definition of chi}
\chi(\lambda)=\sum_n p_{n,N-n}e^{i\lambda n}=\Tr \lp  e^{i\lambda
\hat{N}_A}\rho\rp .
\end{eqnarray}
Since the generating function \eqref{definition of chi} is
represented as an expectation value, $\chi(\lambda)=\la
e^{i\lambda \hat{N}_A}\ra_\rho$, it is often more easy to evaluate
  the quantity \eqref{definition of chi}
than the probability distribution itself.
For fermionic systems, in particular, this can be done with the
help of the determinant representation used to analyze counting
statistics of current fluctuations \cite{LevitovLesovik}.
Once the quantity $\chi(\lambda)$ is known, the probabilities
$p_{n,N-n}$, found from its Fourier transform, can be used to
evaluate the measurement entropy $S_{\rm m}$.
However, if the generating function happens to be Gaussian,
$\chi(\lambda)=e^{-C_2\lambda^2/2}$, the quantity $S_{\rm m}$ can
be evaluated directly as $S_{\rm m}= {1\over 2}\log(2\pi eC_2)$ to
leading order in ${C_2}$.

Below we
 apply this  approach to several cases of interest. We
first consider the correction due to $S_m$ for for entanglement
entropy generated in the process of connecting two initially
separate systems \cite{Klich Levitov}. The entanglement generated
in this case is logarithmic in observation time. In another case,
considered in \cite{Beenakker} entanglement is generated per unit
time in a nonequilibrium, but {\it steady state} of an open
system. Here particles are transmitted across a QPC in the
presence of a bias voltage. In \cite{Beenakker} the role of
super-selection by a stronger constraint, that of energy
conservation, was computed, leading to a correction to
entanglement entropy also scaling linearly with measurement time.
We find that if only number conservation is taken into account,
the correction to accessible entropy is in fact sub-leading.

We then proceed to consider the accessible entropy of critical
systems. Such systems are of particular interest in the theory of
entanglement entropy due to the presence of enhanced, non local,
correlations \cite{VidalLatorreetal,Refael Moore,Wolf,Gioev
Klich,Jin Korepin,Cardy}.
Here we focus on an interacting system in
$1d$, namely a Luttinger liquid, whose entropy can be obtained
using conformal field theory methods. Somewhat surprisingly, the
accessible entropy of this strongly correlated system can also be
found since $S_m$ may be evaluated using bosonization. Finally, we
consider a gapless system
  of free fermions  in higher dimensions.
 We show that, in an arbitrary space dimension $d$, the quantity $S_m$ can be estimated
using Widom's conjecture \cite{Widom82}.
In all of the cases,
$S_m$ turns out to be a
sub-leading contribution to the scaling of many-body entanglement
entropy.

(1)  First, we consider two Fermi seas coupled through a QPC with an externally controled transmission coefficient (Fig.\,\ref{fig1}a).
Initially the two
 Fermi seas  are disconnected and there is
no entanglement between them. The QPC is then opened during the time interval
$0<t<\Delta t$, and then closed again. The  resulting  entanglement entropy
was computed in
\cite{Klich Levitov}.

The measurement entropy of particle number fluctuations the Fermi seas, which gives the correction to entanglement entropy
due to charge conservation, can be evaluated as follows. We use the generating
function $\chi(\lambda)$ defines by \eqref{definition of chi}, where
the state $\rho$ is the density matrix of the two-lead system at
the final time $t=\Delta t$, and $\hat{N}_A=\int_{x<0}
\psi^{\dag}(x)\psi(x)\D x$ is the charge operator in
the left lead.
%
%
The quantity $\chi(\lambda)$ in this situation was analyzed in Ref.\cite{Klich
Levitov} with the help of an expression \cite{LevitovLesovik04}
\begin{eqnarray}
\chi(\lambda)=\exp\lp -{\lambda_*}^2\log{\Delta t\over \tau}\over 2\pi^2\rp
,\quad \sin{\lambda_*\over 2}=\sqrt{D}\sin{\lambda\over 2}
,
\end{eqnarray}
where $D$ is the QPC transmission during $0<t<\Delta t$, and
$\tau$ is a short time
cutoff of the order of the time it takes to switch the QPC from a
disconnected to a connected state. In this case $C_2={D\over
\pi^2}\log{\Delta t\over \tau} $. For large $\Delta t$, the main contribution comes from small
$\lambda$, and the distribution is Gaussian to a good
approximation, in formal analogy to the Central Limit Theorem.

We therefore conclude that $S_{\rm m}$ is given by the entropy of
a Gaussian probability distribution, i.e. $S_{\rm m}\sim {1\over
2}\log2\pi e C_2\sim {1\over 2}\log\log {\Delta t\over \tau}$.
Since the full entanglement entropy generated in this processes is
$S\sim {\pi^2\over 3}C_2$ \cite{Klich Levitov}, we see that the
$S_{res}=S$ to leading order in $C_2$. This is our first concrete example for our general expectation that super-selection leads to a small correction to entanglement entropy.


(2) Next, we analyze entropy generated by a voltage-biased QPC (see Fig.\,\ref{fig1}b).
To leading order in the time of measurement $\Delta t$,
 we can characterize the systems by $N={\Delta teV/h}$
independent transmission attempts,
 with the probability of success equal $D$ for each attempt.  (The
effects of QPC opening and closing are sub-leading since they are
logarithmic in $\Delta t$, as in the previous case.) The
probability to transmit $n$ particles out of $N$
attempts is given by
\begin{eqnarray}
p_n={N!\over (N-n)!n!}D^n(1-D)^{N-n}
\end{eqnarray}
The large-$N$ asymptotic form of the measurement entropy  $S_{\rm
m}=-\sum p_n\log p_n$ of overall transmitted charge is given by
\cite{Jacquet}
\begin{eqnarray}\label{Sm binomial}
S_{\rm m}={1\over 2}\log(2\pi eD(1-D) N),
\end{eqnarray}
whereas the entanglement entropy,  generated in this process is \cite{Beenakker}
\begin{eqnarray}\label{EE biased}
S=N(D\log D+(1-D)\log(1-D))
.
\end{eqnarray}
We see again that $S_{\rm m}$, Eq.\eqref{Sm binomial}, is a sub-leading
correction to the entropy \eqref{EE biased}.

One may further restrict the
allowed operations by not considering entanglement between
particles that cross the barrier with different energies. This
restriction, which was considered in \cite{Beenakker}, substantially
reduces the entropy, since the measurement entropy of energy turns
out to be also linear in $N$.

(3)
Now we shall analyze an interacting fermion systems in one dimension, described by Luttinger liquid model. Below we shall evaluate the correction to the entanglement entropy of a region $A$ of length $L$ due to fermion number
conservation. In bosonization framework, the chiral
Luttinger liquid is a conformal theory with central charge $c=1$.
Therefore its entanglement entropy scales as ${1\over
3}\log(L/a)$, where $a$ is a short-distance cutoff
\cite{Holzhey94}. However, since the Luttinger system is
fermionic,  this entanglement entropy again includes sectors which
mix number of particles. Since the overall charge of the system is
still a good quantum number, the relation \eqref{Sr=S-Sm} holds,
 and so we can estimate the difference $S_A-S_A^{\rm res}$ by evaluating $S_{\rm m}$ for this situation. The Luttinger
Hamiltonian can be written as \cite{KaneFisher}
\[
H={v\over 2}\int\D x :{1\over
g}\partial_x\Theta(x)^2+g\partial_x\phi(x)^2:,
\]
where $\Theta(x)$ and $\partial\phi(x)$ are canonical conjugates
and $g$ depends on the interaction. Within bosonization
$\psi(x)\propto \sum_{n\,{\rm
odd}}e^{-i\sqrt{\pi}\phi(x)}e^{-in(\sqrt{\pi}\Theta(x)+k_Fx)}$.
The density is given by $\rho=\partial_x\Theta(x)+k_F/\pi$ and so
$\hat{N}_A=\Theta(l)-\Theta(0)+k_F L/\pi$.

The generating function \eqref{definition of chi} can now be
computed from the bosonic theory:
\begin{eqnarray}
\chi_{Lutt}(\lambda)=\la e^{i{\lambda\hat{N}_A}}\ra=e^{i\lambda
k_F L/\pi-{g\lambda^2\over 4\pi}\log({k_F L}) },
\end{eqnarray}
giving a Gaussian distribution of charge. Therefore $S_{\rm m}$ for this
case scales as $\log\log k_F L$

\begin{figure}
\includegraphics*[scale=0.4]{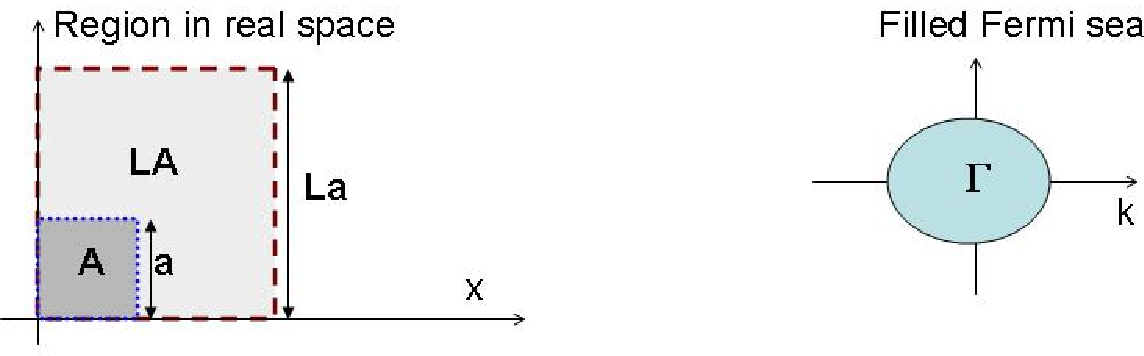}
\caption{The region $LA$, in which particle fluctuations are analyzed, is shown along with the region $A$ of the same shape, having size of order unity. The scaling factor $L$ is used to define scaling of particle number mean and variance, Eq.(\ref{final result widom}).
A region $\Gamma$ in momentum space defines the Fermi
sea, Eq.(\ref{Fermi_distribution})
}
\label{fig2}
\end{figure}

(4) The last case we shall consider is a problem of free fermions in $d$ dimensions. Here the state of the system is described
by a Fermi sea
\begin{eqnarray}\label{Fermi_distribution}
|\psi\ra=\prod_{k\in\Gamma}a^{\dag}_k|0\ra,
\end{eqnarray}
where $\Gamma$ is a domain in momentum space, illustrated in Fig. \ref{fig2},
defining the set of occupied states.
The entanglement entropy of a
region $A$ in real space has been studied in \cite{Wolf,Gioev
Klich} in the asymptotic limit where the linear size of $A$ is
rescaled by a large factor $L$.

To compute the correction due to charge conservation we again
evaluate the measurement entropy $S_{\rm m}$ using
$\chi(\lambda)$.
 The  asymptotic form of the
generating function $\chi(\lambda)$
  at large $L$ can be analyzed as follows.
Let $P_{LA}$ be a projection on the region $A$ in {\it
real} space rescaled by a factor $L$ (i.e. the set of points $Lx$
where $x\in A$,   as illustrated in Fig.\ref{fig2} ). Let
$P_{\Gamma}$ be a projection on the set $\Gamma$ in momentum
space,   obtained from the Fermi distribution \eqref{Fermi_distribution}.
  The generating function $\chi(\lambda)$ can be written as
\cite{Klich03,Budde}
\begin{eqnarray}&
\log\chi(\lambda)=\log\la e^{i{\lambda}\hat{N}_{LA}}\ra=
\Tr\log(1-P_{\Gamma}+P_{\Gamma}e^{i\lambda P_{LA}})
\nonumber \\
& = \Tr\log(1+P_{LA}P_{\Gamma}P_{LA}(e^{i\lambda}-1))
\end{eqnarray}
where $\hat{N}_{LA}=\int_{LA} \psi^{\dag}_x\psi_x\D x$.

We now estimate $\chi$ using Widom's conjecture \cite{Widom82}.
This method has been used in \cite{Gioev Klich} to estimate the
scaling of entanglement of free fermions in arbitrary dimensions.
While a rigorous proof for the conjecture is still missing, it
seems to be perfectly consistent with numerical computations
\cite{Barthel06,Li06}. Widom's conjecture \cite{Widom82} is a
generalization of the strong Szeg\"o theorem to higher dimensions.
It states that given a function $f(z)$, which is analytic on
$|z|\leq1$,
with $f(z)=0$, the following holds as ${L}\to\infty$:
\begin{eqnarray}
\label{Wconj}
  \Tr f(P_{\Gamma}P_{LA}P_{\Gamma})
 =c_1 f(1)L^d+c_2U(f)L^{d-1}\log L
 && \\  \nonumber
+ o({L}^{d-1}\log{L}),\quad
U(f) =\int_0^1 \frac{f(t)-tf(1)}{t(1-t)}\,\D t
.
 &&
\end{eqnarray}
Here the notation $g=o(h)$ means that $g/h\rightarrow 0$ when
$L\rightarrow\infty$, and the coefficients $c_{1,2}$ are given by
$c_1= \frac{1}{(2\pi)^d} \int_A\int_\Gamma
       \D x\D{p}$,
$c_2=\frac{{\log 2}}{(2\pi)^{d+1}}               \,
     \int_{\partial A}\int_{\partial\Gamma}
          |\vec n_x\cdot \vec n_{p}| \D S_x\D S_{p}$,
where $\vec n_x$, $\vec n_{p}$ are unit normals to $\partial A$, $\partial\Gamma$,
respectively.
The formula was proved for $d=1$ in
\cite{LandauWidom80}.

Plugging $f(z)=\log(1+z(e^{i\lambda}-1))$ in \eqref{Wconj}, we
find that $U(f)=- {\lambda^2\over 2}$, and so
\begin{eqnarray}
\label{final result widom}
\log\chi(\lambda)=i\lambda c_1 L^d - {\lambda^2\over 2}c_2 L^{d-1}\log L .
\end{eqnarray}
Thus the charge distribution is Gaussian to leading order, and
$S_{\rm m}$ scales as $\log (L^{d-1}\log L)$ and is { again} a sub-leading
correction to the entropy.

In summary, particle number and charge conservation, as well as
spin conservation, under some conditions, is
 an essential part of realistic Fermi systems.
However, the existing discussions of many body entanglement of
fermions \cite{Wolf,Gioev Klich,Klich Levitov} implicitly assume
that such conservation laws have no direct effect on the scaling
of many-body entanglement. The analysis of the measurement entropy
of particle number fluctuations and of its relation to
super-selection rules and entanglement entropy, presented above,
indicates that this expectation is in fact correct.
We analyzed several systems of interest, including time dependent
scattering problems, the ground state of one-dimensional
interacting fermions (a Luttinger liquid), and a free fermion
system in higher dimensions. In all those cases we found that
super-selection rules yield sub-leading corrections to the
entanglement entropy. We conclude that for a generic Fermi system
in dimension $d>1$ the accessible entropy is equal, to leading
order in system size, to the full entanglement entropy. The same
is true for critical systems in $d=1$ described by conformal field
theories with density correlator decaying as $1/r^2$.


We would like to thank F. Verstraete and C. W. J. Beenakker for
useful discussions. We acknowledge support by the National Science
Foundation under Grant No. PHY05-51164. The work of L. L. was
supported in part by W. M. Keck Foundation Center for Extreme
Quantum Information Theory.

\end{document}